\documentclass[conference]{IEEEtran}
\IEEEoverridecommandlockouts
\usepackage{cite}
\usepackage{amsmath,amssymb,amsfonts,times,graphicx,booktabs,tabulary,multirow,overpic,xcolor,url}
\usepackage{algorithmic}
\usepackage{graphicx}
\usepackage{textcomp}
\usepackage{xcolor}
\usepackage[tableposition=top]{caption}
\def\BibTeX{{\rm B\kern-.05em{\sc i\kern-.025em b}\kern-.08em
    T\kern-.1667em\lower.7ex\hbox{E}\kern-.125emX}}
\begin{document}

\title{Financial Markets Prediction with Deep Learning}

\author{\IEEEauthorblockN{Jia Wang\IEEEauthorrefmark{1}, Tong Sun\IEEEauthorrefmark{1}, Benyuan Liu\IEEEauthorrefmark{1}, Yu Cao\IEEEauthorrefmark{1} and Degang Wang\IEEEauthorrefmark{2}}
\IEEEauthorblockA{\IEEEauthorrefmark{1}Department of Computer Science\\
University of Massachusetts Lowell, Lowell, Massachusetts 01854\\
Email: \{jwang,tsun,bliu,ycao\}@cs.uml.edu}
\IEEEauthorblockA{\IEEEauthorrefmark{2}GRC Capital LLC, Wellesley, Massachusetts 02481\\
Email: degang.wang@gmail.com}}

\maketitle

\begin{abstract}
Financial markets are difficult to predict due to its complex systems dynamics. Although there have been some recent studies that use machine learning techniques for financial markets prediction, they do not offer satisfactory performance on financial returns. We propose a novel one-dimensional convolutional neural networks (CNN) model to predict financial market movement. The customized one-dimensional convolutional layers scan financial trading data through time, while different types of data, such as prices and volume, share parameters (kernels) with each other. Our model automatically extracts features instead of using traditional technical indicators and thus can avoid biases caused by selection of technical indicators and pre-defined coefficients in technical indicators. We evaluate the performance of our prediction model with strictly backtesting on historical trading data of six futures from January 2010 to October 2017. The experiment results show that our CNN model can effectively extract more generalized and informative features than traditional technical indicators, and achieves more robust and profitable financial performance than previous machine learning approaches. 
\end{abstract}


\section{Introduction}\label{Introduction}

Financial market is a very complex adaptive system. The complexity mainly derives from the interaction among markets and market participants---the current environment of markets influence strategies of market participants, while the overall behavior of market participants decides the trend of financial market. According to the Adaptive Markets Hypothesis (AMH)~\cite{b17}, behavioral biases of market participants, such as loss aversion, overconfidence, and overreaction, always exist. Moreover, once market environment changes, the heuristics of the old environment may not work any longer and thus degenerate to behavioral biases. As a result, financial market's trend interwoven with market participants' biased strategies make financial market very difficult to predict.

There have been many attempts to predict the market movement using various approaches. Technical indicators/technical analysis~\cite{b1} has been traditionally used in ~\cite{b3,b4,b24,b2}, but these methods tend to lose predictive power after they are published. Recently there have been some studies that use machine learning techniques (e.g., feedforward neural networks, SVM, ensembles) for financial market prediction~\cite{b22,b6}. They achieved good performance in terms of the benchmark of machine learning, however, have no comparable results from the financial perspective.

In a nutshell, the essence of financial markets prediction is to find out generalized and informative features from the joint distribution of prices and volumes~\cite{b16}. We need to train a model to represent a generalized joint distribution based on current market environment. In this paper, we propose and develop a deep convolutional neural network model (CNN) to automatically extract features from historical financial trading data and to predict the price movement. Through the multi-layer customized 1-D convolutions, noise can be filtered out, meanwhile highly correlated underlying features emerge and are clustered to corresponding channels for further feature combinations in fully connected layers. The whole process is end-to-end so that potentially negative influence brought by human interference, such as selection of technical indicators and pre-defined coefficients in technical indicators, can be avoided. 

To our knowledge, this is among the first efforts to use deep CNN to predict financial markets and we verify the performance of our model using strict backtests. We test our model with six futures from Chicago Mercantile Exchange and New York Mercantile Exchange. Backtest results show that our 1-D CNN model achieves significantly higher average annual return and more robust performance (higher Sharpe ratio\footnote{Sharpe ratio is the average return earned in excess of the risk-free rate per unit of volatility or total risk. It is widely used for calculating risk-adjusted return.}) over previous approaches based on Nearest Neighbor, SVM, and Deep Feedforward Networks.

We also observe that our 1-D CNN model without using the technical indicators as input achieves better results than the model that uses the technical indicators. This shows that our 1-D CNN model can effectively extract more generalized and informative features than those represented by traditional technical indicators. Our results confirm the observation that common metrics in machine learning, such as accuracy and F1 score, are not suitable for financial markets prediction because different types of prediction errors have different impacts on financial performance~\cite{b22}. In our study, we propose a modified version of the F-measure score, called Weighted-F-Score, to address this issue. Our experiments show that Weighted-F-Score highly correlates with average annual return and Sharpe ratio in our backtest results with the minimum cross-correlate coefficient of 0.79 and 0.84, respectively.

We summarize our contributions as follows:
\begin{enumerate}
\item \textbf{Cross-Data-Type One-dimensional Convolution.} Regular 2-D convolutions are not directly applicable for this research because we may not convolve different elements of financial historical trading data, such as price and volume, to produce meaningful results. The regular 1-D convolution do not apply either because it can not capture the features to represent the joint distribution of elements of financial historical trading data. To address the above problem, we propose a variant of 1-D convolution, called Cross-Data-Type 1-D convolution. Based on the customized 1-D convolution, our model outperforms other methods by 6.1\%-53.0\% on average annual return and 53.0\%-199.0\% on Sharpe ratio.
\item \textbf{Deep Features Auto-extraction.} Many machine learning methods for financial market prediction use technical indicators as input. We propose to replace traditional technical indicators with deep features extracted by Cross-Data-Type 1-D convolution in order to capture more generalized and informative features. To our best knowledge, this is among the first methods that use CNN to extract features instead of traditional technical indicators. Our experiments show this method achieves state-of-the-art result both in finance return and machine learning benchmark.
\item \textbf{Correlation between Finance and Machine Learning metrics.} Different types of predicting errors have different impacts on financial market trading. For example, the error that the prediction is {\em up} when the price actually goes down or vice versa leads to more loss in trading. However, common metrics of machine learning, such as accuracy and F1 score, do not discriminate different types of errors so that they weakly correlate with financial metrics. We propose a modified version of F score for Type I and Type II errors, called Weighted-F-Score, in order to address the weak correlation between Machine Learning and Finance metrics. Our results show that Weighted-F-Score highly correlates with Finance metrics (the minimum Cross-correlation value is 0.79).
\end{enumerate}

The remainder of the paper is structured as follows. Related work on financial market prediction with technical analysis and machine learning methods is presented in Section \ref{Related Work}. The architecture and customized convolutional kernel design of our 1-D CNN model are presented in Section \ref{CNN}. The datasets, the backtest strategy, and the machine learning metric that we use in this research, Weighted-F-Score, are described in Section \ref{Experimental setup}. The experiments results are presented in Section \ref{Experiments and Results}, followed by discussing and concluding remarks.

\section{Related Work}\label{Related Work}

As a method to predict price movements of financial markets, technical indicators/technical analysis has been extensively used by market practitioners since 1800s \cite{b21}. Although it is still controversial whether technical analysis goes against Efficient Market Hypothesis~\cite{b23,b20}, some studies, such as \cite{b15} and \cite{b26}, provide evidences of its forecastability.

Following traditional charting analysis, researchers' interest has turned into automated trading based on pattern recognition and machine learning since the 1990s.  \cite{b8} proposes to predict one-year daily long or short signals by a single-layer feedforward networks trained by last four-year daily returns of Dow Jones Industrial Average. \cite{b7} proposes a nonlinear model based on technical indicators model that outperforms moving average rules in terms of both returns and Sharpe ratio in European exchange markets for 1978-1994. \cite{b13} applies support vector machines to the daily Korea composite stock price index with 12 technical indicators. \cite{b9} introduces recurrent neural networks to address difficulties with non-stationarity, overfitting, and unequal a priori class probabilities. \cite{b22} introduces ensembles to improve forecastability based on decision trees, SVM, and feedforward networks.

Deep Learning has been popular in many fields of artificial intelligence since the breakthrough on Imagenet challenge by \cite{b14}. \cite{b6} applies 5-layer Deep Feedforward Networks to predict 43 Commodity and FX futures prices' trend. \cite{b5} proposes a deep learning method to predict stock markets using Natural Language Processing on financial news. \cite{b28} uses Long Short Term Memory networks \cite{b29} combined with Inverse Fourier Transform to capture multiple frequency features underlying the fluctuation
of stock prices.

It is worth noting that among the aforementioned studies for automated trading, few of them has released related results about financial profitability. To our knowledge, \cite{b6,b7} presents a series of financial based results, such as cumulative return and Sharpe ratio, although the results are not very encouraging and their models' profitability may not be convincing. 

\section{Model Design}\label{CNN}

\subsection{Cross-Data-Type 1-D CNN}\label{Basic Architecture}

\begin{figure*}
\centering
\includegraphics[width=1\linewidth]{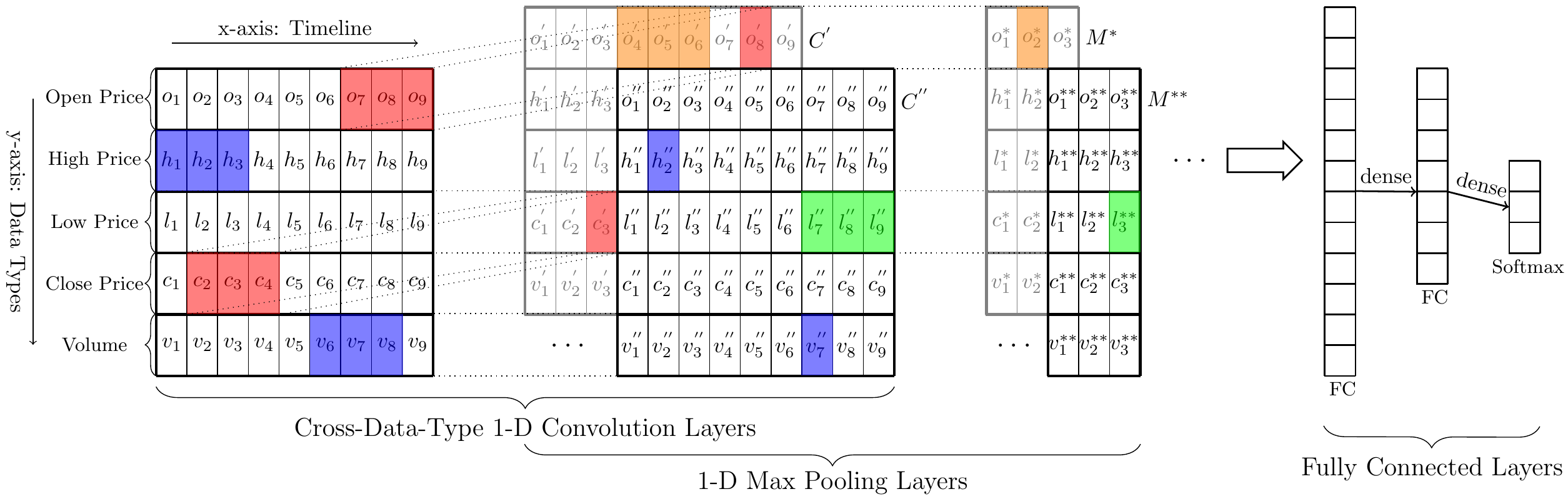}
\caption{\textbf{Cross-Data-Type 1-D Convolution Architecture}. The 1-D kernels (the red and blue one) scan along with the x-axis while each one of them goes to every position of input 2-D frames by a stride of one. The $C^{'}$ and $C^{''}$ represent the output channels of the red and blue kernel, respectively. A max-pooling layers follows a convolutional layer, and only max-pooling layers condense the dimensions of the x-axis. The $M^{*}$ and $M^{**}$ represent the output channels of the max-pooling operations (the orange and green one) by a stride of three.}
\label{fig:cnn}
\end{figure*}

CNN has been known as an effective model for automatic feature extraction in many fields such as computer vision and natural language processing. However, the structure of financial data does not directly allow for regular convolutions. On one hand, although we can organize financial trading records into 2-D frames, the regular 2-D convolution is not suitable because convolving different types of data may produce nonsensical results. On the other hand, the regular 1-D convolution is defective because data of each type needs parameter sharing while it also needs clear boundaries in order to avoid producing nonsensical results as in the regular 2-D convolution case. To address this problem, we propose a novel 1-D convolution architecture called Cross-Data-Type 1-D CNN (CDT 1-D CNN). As illustrated in Figure \ref{fig:cnn}, input data is organized into 2-D frames, where the x-axis denotes time and contains a certain number of trading records, and the y-axis denotes data types and contains prices, volumes, and technical indicators (if any). All kernels are one-dimensional and only scan along the x-axis. After each convolutional layer, a max-pooling layer is used to reduce the x-axis dimension of feature maps. 

The scan strategy of the CDT 1-D kernels is different from that of the regular 1-D or 2-D convolution: all the kernels scan only along the x-axis while each one of them goes to every position of the 2-D frames. That means, once a kernel finishes the scan for one row (the x-axis), it will turn to scan from the start point of the next row and so forth until the whole 2-D frame is scanned. As illustrated in Figure \ref{fig:cnn}, the 1x3 red and blue kernel scan the whole 2-D input frame respectively, and they only touch three elements of one data type at each time. In other words, data of different types can share parameters by the scan strategy. It is crucial because it matches up the characteristics of the structure of financial data. Unlike in computer vision's datasets, the 2-D frames contain only one type of data {\em pixel},  the 2-D frames in our datasets contains prices, volumes, and technical indicators (if any). However, according to ~\cite{b16}, the joint distribution of these types of data has strong forecastability over marginal distributions of every single one. Therefore, on one hand, we cannot directly convolve different types of data; on the other hand, we want to preserve the parameter sharing so that the model can represent the desirable joint distribution effectively. Although this is evident in computer vision applications, in our study, it needs more explanation. Take one technical indicator, Typical Point ($TP$), for instance,
\begin{equation}\label{eq:tp}
TP = \frac{P_{high} + P_{low} + P_{close}}{3}
\end{equation}
Now assume that these three prices need to be scaled for training purposes and assume that the effect of the activation is ignored, we can modify Equation (\ref{eq:tp}) to:
\begin{equation}
TP^{'} = \frac{P_{high}^{'} + P_{low}^{'} + P_{close}^{'}}{3}
\end{equation}
where $P_{high}^{'}=\alpha_{high}P_{high}$, $P_{low}^{'}=\alpha_{low}P_{low}$ , $P_{close}^{'}=\alpha_{close}P_{close}$. To preserve the properties of $TP$ after the scaling, we should have $\alpha_{high}=\alpha_{low}=\alpha_{close}$. Note that the parameter sharing in the CDT 1-D CNN guarantees outputs from the same channel to be scaled by the same value. In this mutual relationship among different data types are preserved and inherited layer after layer, and thus underlying features emerge with training iterations.

\subsection{Without technical indicators}\label{Without technical indicators}
Technical indicators may capture profitable patterns in financial markets, which is the main reason why it has been popular among market practitioners over one hundred years. However, how to choose indicators relies on the experience of practitioners and the popularity of indicators. Although feature selection methods such as Information Gain and correlation-based filters \cite{b22} can be helpful, the indicator candidates pool still limits the model's performance. Also, some machine learning algorithms with technical indicators, such as Nearest-neighbour and SVM, are reported to achieve occasional rather than consistent success \cite{b26}. These facts may reveal the limitation of technical indicators/technical analysis that human interventions, such as choosing technical indicators and defining parameters of technical indicators, bring negative effects. Inspired by the evolution of computer vision research from extracting hand-designed features such as SIFT \cite{b19} to autonomously learning features within training process \cite{b14,b27}, we argue that the CDT 1-D CNN model is more suitable for profitable patterns extraction over technical analysis by the three following reasons:
\begin{enumerate}
\item The CDT 1-D CNN model itself is a nonlinear and autonomously-trained variant of technical analysis. The fundamental elements of technical analysis, technical indicators, which are mathematical formulas based on historical data of prices and volumes, may be approximated by the convolutional and fully connected layers. Take another technical indicators, Moving Average Convergence Divergence (MACD), for instance:
\begin{equation}\label{eq:ema}
EMA_{n}[i]=\alpha P_{close}[i] + (1 - \alpha) EMA_{n}[i-1]
\end{equation}
\begin{equation}\label{eq:macd}
MACD=EMA_{a} - EMA_{b}
\end{equation}
where $n$, $a$, and $b$ denote time spans, and $a$ $>$ $b$. $i$ denotes the $i_{th}$ time of a given time span, and $i \in [1,n]$. $P_{close}[i]$ denotes the close price at the $i_{th}$ time, $EMA_{n}[i]$ denotes Exponential Moving Average, and $EMA_{n}[1]=P_{close}[1]$. $\alpha$ denotes a predefined coefficient. $EMA_{n}$ is, in fact, equivalent to an infinite impulse response (IIR) filter for $P_{close}$ in the period $n$, which can be implemented by convolutions. Thus after a subtraction operation provided in fully connected layers, the functionality of $MACD$ can be realized.
\item More importantly, in previous studies the coefficients of technical indicators, such as $\alpha$, $n$, $a$, and $b$ in Equation (\ref{eq:ema}) and (\ref{eq:macd}), are predefined by users and thus not trainable. However, deep learning models need to train parameters to learn distributed representations with learning algorithms, such as loss function and back propagation. Therefore, non-trainable coefficients of technical indicators are inflexible for training process so that the effectiveness of technical indicators relies on user experience and knowledge rather than the joint distribution of training data. Formally, our model computes a function as follows: 
\begin{equation}
y_{t}^{'} = F(\vec{X_{t}}, W)
\end{equation}
where $X_{t}$ is the $t$-th input vector, and $W$ represents trainable parameters (e.g. weights of convolution kernels and fully connected layers) in the model. Meanwhile, the model minimizes a loss function within the $t$-th prediction $y_{t}^{'}$ and the $t$-th ground truth $y_{t}$: 
\begin{equation}
E_{t} = L(y_{t}^{'}, y_{t})
\end{equation}
In other words, the training process is to find the value of $W$ that minimizes $E_{t}$. Now assume the $\vec{X_{t}}$ is a vector of technical indicators at the $t$-th time as follows:
\begin{equation}
\vec{X_{t}} = \bigcup_{a \in N}{T_{a}(\vec{x_{t}}, \Theta)}
\end{equation}
where $\vec{x_{t}}$ is the raw financial trading data vector at the $t$-th time that includes prices and volume only, $\bigcup_{a \in N}{T()}$ represents a set of functions of technical indicators, and $\Theta$ denotes a collection of predefined coefficients in the functions of technical indicators. Since prices and volume in raw financial trading data reflect all available information \cite{b20}, and all technical indicators are based on raw financial trading data. Therefore, technical indicators do not provide additional information beyond raw financial trading data, and prediction models cannot adjust predefined $\Theta$ so that bias and overfitting issues may occur.

\end{enumerate}


\section{Experimental setup}\label{Experimental setup}
\subsection{Data}\label{Data}

\begin{figure*}
\centering
\includegraphics[width=1\linewidth]{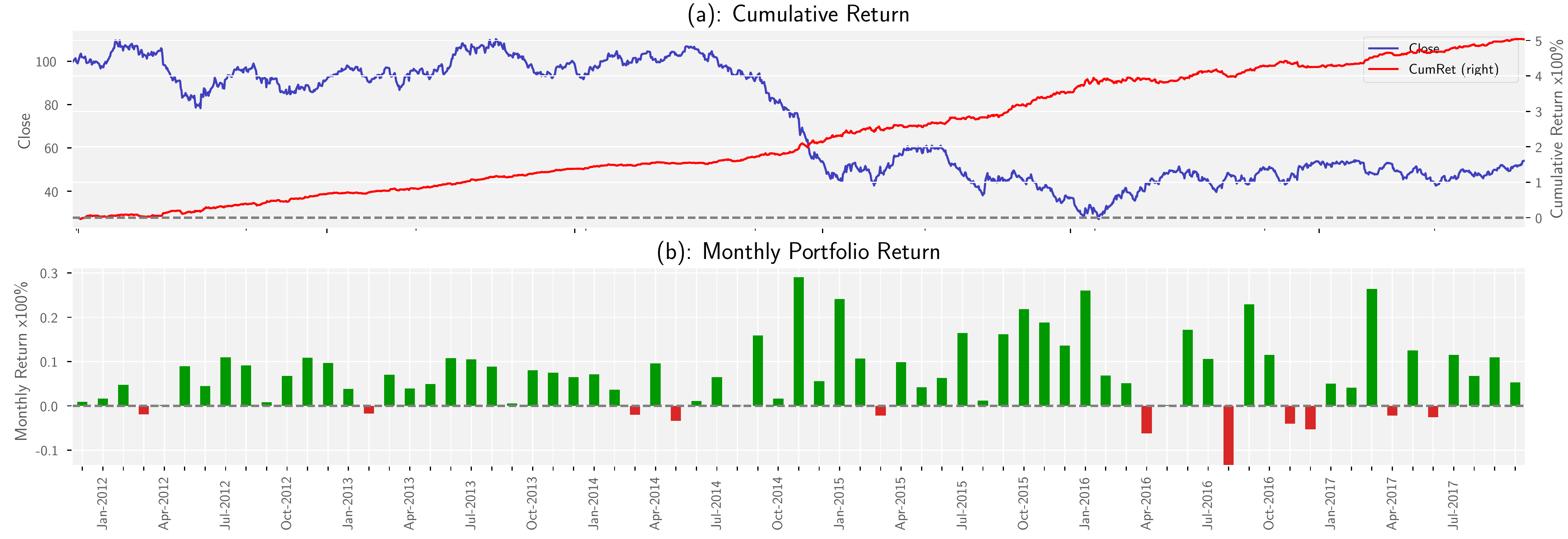}
\caption{\textbf{Cumulative and Monthly Return of the future CL by {\em the CDT 1-D CNN without technical indicators} }. (a): The cumulative return of the future CL stays positive over 71 months, and it eventually achieves 500\%. (b): The Monthly Return of the future CL over 71 months. 60 of the 71 months are profitable, and only two months have negative return lower than -5\%. }
\label{fig:cm}
\end{figure*}

The datasets we used in our study are historical trading records of four commodity futures and two equity index futures, including WTI Crude Oil (CL), Natural Gas (NG), Soybeans (S), Gold (GC), E-mini Nasdaq 100 (NQ), and E-mini S\&P 500 (ES). Since there are no open datasets available thus far, the six datasets in our study are collected from online brokers. The datasets of the six futures range from January 2010 to October 2017, each dataset contains 330,000-400,000 5-minute trading records. Each record has the following seven attributes: date, time, open price, high price, low price, close price, and trading volume. 

A large variety of technical indicators have been developed to predict the future price levels, or simply the general price direction, of a security by looking at past patterns. We choose technical indicators according to their functionality and popularity. For lagging indicators, which are often used to identify and confirm the strength of a pattern or trend, we pick EMA, MACD Histogram, and Bollinger Bands. For leading indicators, which usually change before a trend or pattern and are thus used during periods of sideways or non-trending ranges, we pick Relative Strength Index (RSI),  Commodity Channel Index (CCI), Volume Weighted Average Price (VWAP), On-balance Volume (OBV), Average Directional Index (ADX), Accumulation Distribution Line (ADL), and Chaikin Money Flow (CMF). Also, we include technical indicators regarded as neither lagging nor leading indicators such as Rate of Change (ROC), which is used to measure the change of prices over time. Different time spans are used to generate EMA, MACD Histogram, Bollinger Bands, ROC, RSI, CCI, CMF, and ADX. Together with the original five attributes (open, high, low, close prices, and trading volumes), each data point has a total number of 46 attributes.

Our model predicts the future price movement every two hours and makes the trading decisions accordingly based on the prediction outcome. The reason that our model works with a two-hour time interval is two folds:
\begin{enumerate}
	\item Longer time intervals provide necessarily data sequence size for convolutions. Each 24 consecutive 5-minute data records are organized to form a data frame, where 1-D convolution is applied along the timeline. For the sake of fair comparisons between our MLP and 1-D CNN models, the datasets fed to the MLP model also have the same time interval.
	\item The price change of a commodity over a shorter time interval is usually much smaller than that over a longer time interval. For example, the average price change of CL over a 5-minute time interval is \$0.02 compared to \$0.10 over a 2-hour time interval. The profitability of trading over short time intervals is thus diminished after slippage and transaction cost are applied. 
\end{enumerate}

Another vital part of our data pre-processing is labeling. The popular way for labeling stock or future price movements is based on a fixed threshold for the log return of the price. For example, the data point at time $t$ is marked as ``going up" or ``going down" if the log return of the price at time $t+1$ over $t$ is above or below the pre-specified threshold. However, because the aforementioned method does not thoroughly consider the distribution of future price's trend \cite{b18}, we adopt the dynamic threshold method as described in \cite{b25}:

\begin{equation}
l_{t+1} =
\begin{cases}
\text{up,}      & \text{if $c_{t+1} \geq c_{t} \left( 1 + \alpha v_{t}\right)$} \\
\text{down,}  & \text{if $c_{t+1} \leq c_{t} \left( 1 - \alpha v_{t}\right)$} \\
\text{flat,}     & \text{otherwise}
\end{cases}
\end{equation}
where $c_{t}$ and $v_{t}$ represent the close price and volatility of a commodity price at time $t$, and $\alpha$ is a parameter that can be adjusted to balance the three classes. In our study, the volatility is measured by using the standard deviation of the past ten data points of any given time and $\alpha$ is set to be 0.55. 

\subsection{Criterion}\label{Criterion}

\subsubsection{Backtest (From Finance Viewpoint)}\label{Backtest}
Previous studies usually choose buy-and-hold strategy as the baseline. However, we believe that this strategy is too simple to serve as a baseline. Take the future CL for instance, if we placed and held one contract since July 2014, our asset thus far would shrink around 70\%. In our research, the backtest strategy that we use is more conservative than that in \cite{b6,b7}. Specifically, we use high transaction cost to make our backtest result more conservative:




\begin{equation}\label{eq:s}
S = 5F
\end{equation}
\begin{equation}\label{eq:b}
B^{'} = 2B
\end{equation}
\begin{equation}\label{eq:t}
T = n (M(B^{'}+S)+C)
\end{equation}
where $S$ denotes slippage, the difference between the expected price of a trade and the price at which the trade is actually executed. $F$ denotes minimum price fluctuation, the smallest increment of price movement possible in trading a given contract. $B$ denotes bid-ask spread, the amount by which the ask price exceeds the bid price for an asset in the market. $B^{'}$ denotes magnified bid-ask spread. $M$ denotes the multiplier, the deliverable quantity of commodities and option contracts that are traded on an exchange. $C$ denotes commissions per contract, the fixed fee to brokers. $n$ denotes the number of contracts. $T$ denotes transaction cost. Note that we intentionally raise $S$ by three times and $B$ by two times to make our backtest result conservative enough. Take CL for instance, the bid-ask spread is \$0.01, the minimum price fluctuation is \$0.01, the multiplier is 1,000 times, and the commissions is \$2.75 at Interactive Brokers\footnote{https://www.interactivebrokers.com}, the profit of a trade with one contract needs to gain no less than \$145.50; otherwise, this trade will end up losing money.

The backtest strategy that we use follows a principle with minimum human interventions---at any time, depending on the predicting price movement at next time interval, the trading strategy determines whether to enter or leave short/long trades.
The detailed breakdown of this strategy is that the first $Up$ or $Down$ predicting label renders the test into the $long$ $trade$ or the $short$ $trade$, respectively. After that, once a turning point appears, for example, the current trading status is the $long$ $trade$ but the next predicting label is $Down$, the trading strategy will leave the $long$ $trade$ and enter the $short$ $trade$. Note that the term $T$ in Equation (\ref{eq:s}) is calculated for each trade. In our experiments, we allocate \$100,000 as initial capital, and use Equation (\ref{eq:s}) for transaction cost. The number of contracts is one, and all other terms in Equation (\ref{eq:s}), (\ref{eq:b}), and (\ref{eq:t}) follow the standard settings of futures markets.

\subsubsection{Weighted F Score (From Machine Learning Viewpoint)}\label{Weighted F Scores}
The commonly used metrics in machine learning, such as accuracy and F1 score, do not correlate well with the trading performance (e.g., profit return, Sharpe ratio) in this research. As shown in our backtest strategy, the $Up$ and $Down$ prediction of commodity prices play important roles in the automated trading process, directly triggering long and short trades and affecting the bottom line.  
We categorize wrong predictions of price movement into three types, in the order of criticality for profitable operations. The first type of errors has a prediction in the opposite direction of actual price movement, i.e., the prediction is "up" when the price goes down, and vice versa. This will lead to trading in the wrong direction and hurt the return.  The second type of errors occurs when the price change is flat, but the model predicts it to go up or down. This will trigger the strategy to enter long or short positions without a clear price moving trend, and may incur a moderate loss caused by slippage and transaction costs. The third type of errors happens when the price moves up or down, but the model predicts it to stay flat, resulting in lost opportunities to enter potentially profitable trades. 


To address the specific characteristics, we propose a modified version of F score, called Weighted F Score. It is similar to the original F score for Type I and Type II errors, however, we use the aforementioned three types of prediction errors instead of Type I and Type II errors. The basic idea of Weighted F Score is that, we assign different weights to different types of error in order to raise the priority of critical errors (e.g., the first type of errors mentioned above) and to lower the priority of noncritical errors (e.g., the third type of errors). Formally, 
\begin{equation}
N_{tp} = N_{tu} + N_{td} + \beta_{3}^2N_{tf}
\end{equation}
\begin{equation}
E_{1st} = E_{wutd} + E_{wdtu}
\end{equation}
\begin{equation}
E_{2nd} = E_{wutf} + E_{wdtf}
\end{equation}
\begin{equation}
E_{3rd} = E_{wftu} + E_{wftd}
\end{equation}
\begin{equation}\label{eqn:FS}
F = \frac{\left(1+\beta_{1}^2+\beta_{2}^2\right)N_{tp}}{\left(1+\beta_{1}^2+\beta_{2}^2\right)N_{tp}+E_{1st}+\beta_{1}^2E_{2nd}+\beta_{2}^2E_{3rd}}
\end{equation}
where $N_{tp}$ denotes True Positive, $E_{1st}, E_{2nd},$ and $E_{3rd}$ denote the first type, second type, and third type of errors, respectively. Moreover,  $N_{tu}$ denotes $True$ $Up$, $N_{td}$ denotes $True$ $Down$, $N_{tf}$ denotes $True$ $Flat$, $E_{wutd}$ denotes $Wrong$ $Up$ $but$ $True$ $Down$, $E_{wdtu}$ denotes $Wrong$ $Down$ $but$ $True$ $Up$, $E_{wutf}$ denotes $Wrong$ $Up$ $but$ $True$ $Flat$, $E_{wdtf}$ denotes $Wrong$ $Down$ $but$ $True$ $Flat$, $E_{wftu}$ denotes $Wrong$ $Flat$ $but$ $True$ $Up$, $E_{wftd}$ denotes $Wrong$ $Flat$ $but$ $True$ $Down$. $\beta_{1}$, $\beta_{2}$, and $\beta_{3}$ denote the weighting parameters for Second-Level, Third-Level Errors, and $True$ $Flat$, respectively.  In our study, we set $\beta_{1}$ to 0.5, $\beta_{2}$ and $\beta_{3}$ to 0.125.

\section{Experiments and Results}\label{Experiments and Results}

We choose Nearest-neighbour, Support Vector Machines (SVM), and Deep Feedforward Networks (MLP) as the baselines~\cite{b7,b13,b6}. The implementation of the MLP and CNN models is based on Tensorflow framework\footnote{https://www.tensorflow.org}, and the implementation of SVM and NN is based on Sklearn framework\footnote{http://scikit-learn.org}. For SVM, we follow the settings by \cite{b13}---the kernel is radial basis function (RBF), $C$ $=$ $78$, and $\sigma^2$ $=$ $25$. Since \cite{b6} does not release the full set of hyperparameters, we use grid search method to define our MLP model's setting---it has 6 hidden layers, and with the ascending order of depth, unit number of each layer is 300, 200, 100, 50, 25, and 10, respectively. Also, we use Adam Optimization Algorithm \cite{b30}, 1E-3 initial training rate, 0.7 dropout, 1E-5 L2 weight decay and 12-sample mini-batch for better generalization \cite{b12}. We also use the rectified linear unit (ReLU) as the activation function and use normalized initialization to initialize weight parameters \cite{b10,b11}. These definitions of hyperparameters are also applied in all the MLP and CNN models. 

\begin{table*}[!t]

\centering
\scalebox{1.2}{
\begin{tabular}{ |*{8}{c|}}
\hline
\multirow{2}{*}{}  & \tiny w/ TIs? & CL & NG & NQ & S & ES & GC  \\
\cline{3-8}
\hline
\emph{\tiny SVM}
  & \checkmark & 50.2\% & 41.0\%  & 45.4\%  & 43.0\%& 41.1\%  & 38.6\% \\
\hline

\emph{\tiny MLP} 
 &  \checkmark & 49.3\%  & 51.1\%  & 52.3\% & 48.1\% & 43.1\% & 41.9\%\\
 \hline

 \emph{\tiny regular 1-D CNN}
 & & 50.4\%  & 50.9\% & 54.8\%  & 50.4\% & 50.8\% & 44.8\% \\
 \hline

 \multirow{2}{*}\emph{\tiny CDT 1-D CNN}
 & & \textbf{57.4\%}  & \textbf{52.5\%}  &  \textbf{55.9\%} & \textbf{51.7\%}  & \textbf{51.8\%}  &\textbf{47.3\%} \\
 \cline{2-8}
 &  \checkmark & 54.1\%  & 48.6\% & 49.2\%  & 49.4\% & 47.7\% & 44.2\% \\
 \hline
\end{tabular}
}
\caption{\textbf{Weighted F Score (WFS)}. {\em The CDT 1-D CNN without technical indicators (TIs)} outperforms other models.}
\label{fr}
\end{table*}

Moreover, sequential training sessions proceed by moving windows. Each moving window consists of a training set, a validation set, and a test set. The size of these sets is 142,416 (2-year), 192 (4-week), and 96 (2-week) trading records, respectively. The three types of datasets are in chronological order to avoid the {\em leaking} problem \cite{b22}. Each training session starts from scratch with the corresponding moving window, and the moving step between adjacent windows is 96 (2-week). The combined prediction results from the whole training sessions are used as input of the backtest.

The CDT 1-D CNN model used in our study consists of three CDT 1-D convolutional and max-pooling layers, two fully connected layers (1,000 and 500 units), and an output softmax layer. With the ascending order of depth, the kernels of each convolutional layer are 4x32, 3x64, and 2x128, respectively. The strides of each max-pooling layer are 4, 3, and 2, respectively. As a result of the CDT 1-D convolutions and max-pooling operations along with x-axis, the dimensions of time decrease from 24 to 1. Meanwhile the dimensions of the channels increase from 1 to 128. 

\begin{figure}
\centering
\includegraphics[width=1\linewidth]{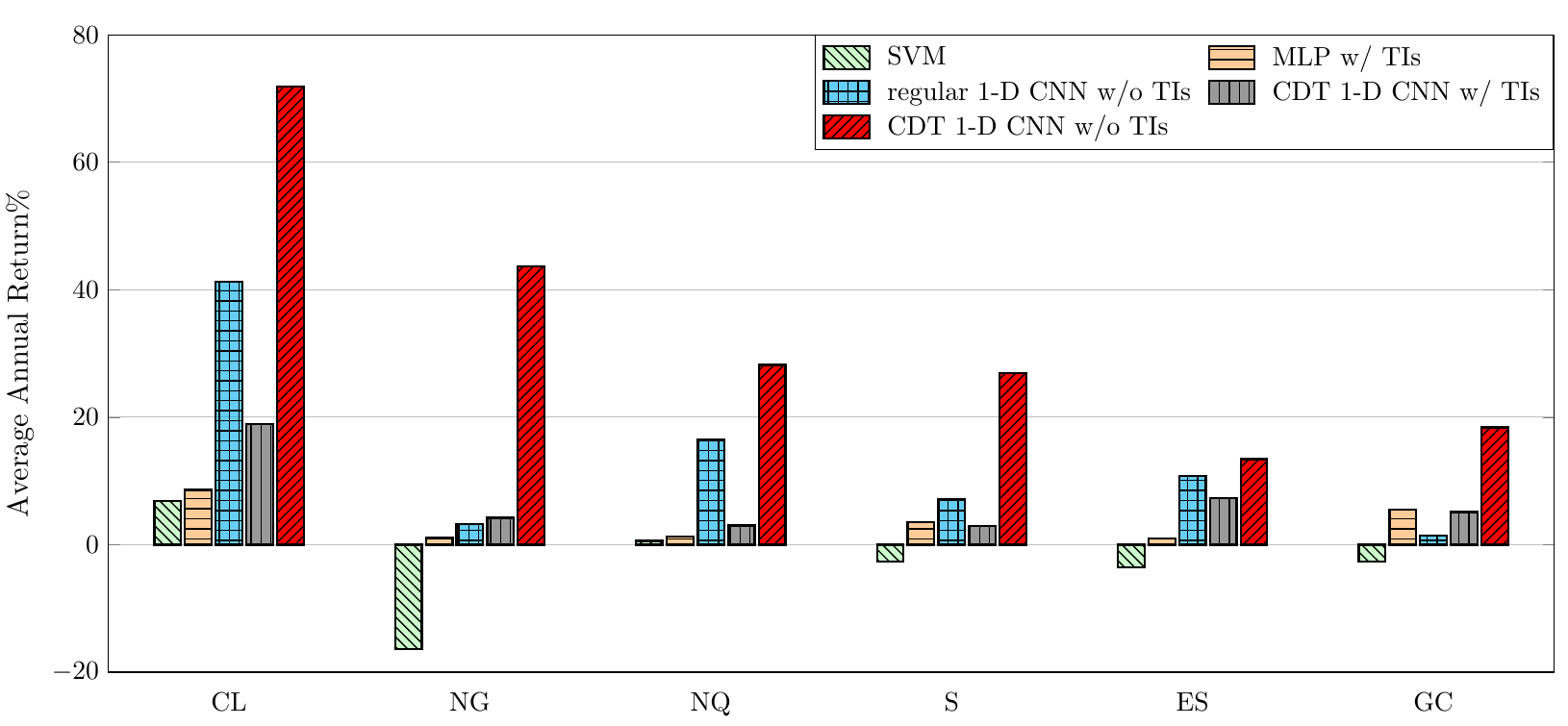}
\caption{\textbf{The average annual return of the six futures over 71 months}. {\em The CDT 1-D CNN without technical indicators (TIs)} outperform the baselines by 12.4\%-63.3\%.}
\label{fig:ar}
\end{figure}

\begin{figure}
\centering
\includegraphics[width=1\linewidth]{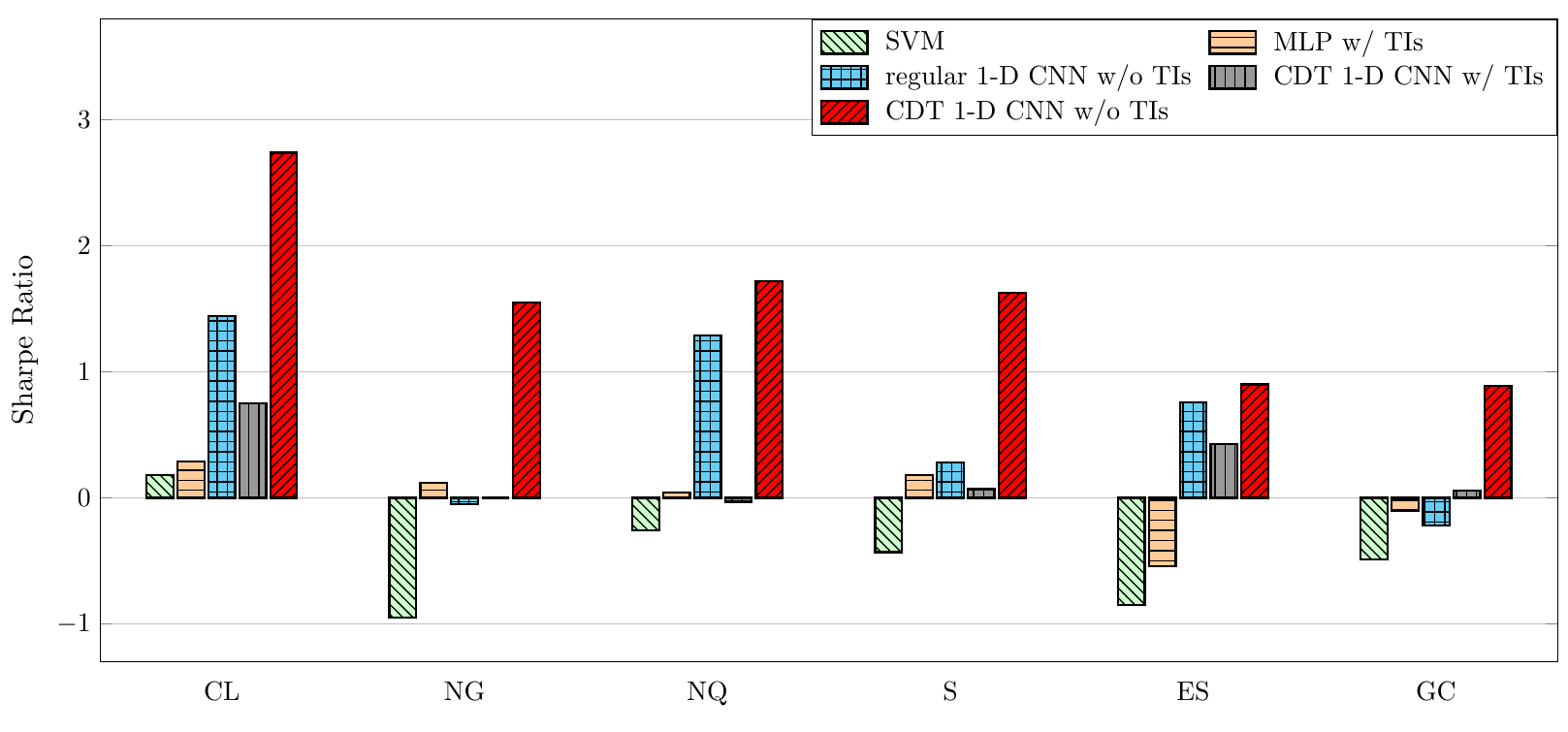}
\caption{\textbf{The Sharpe ratio of the six futures over 71 months}. {\em The CDT 1-D CNN without technical indicators (TIs)} significantly outperform the baselines.}
\label{fig:sr}
\end{figure}

\begin{figure}
\centering
\includegraphics[width=1\linewidth]{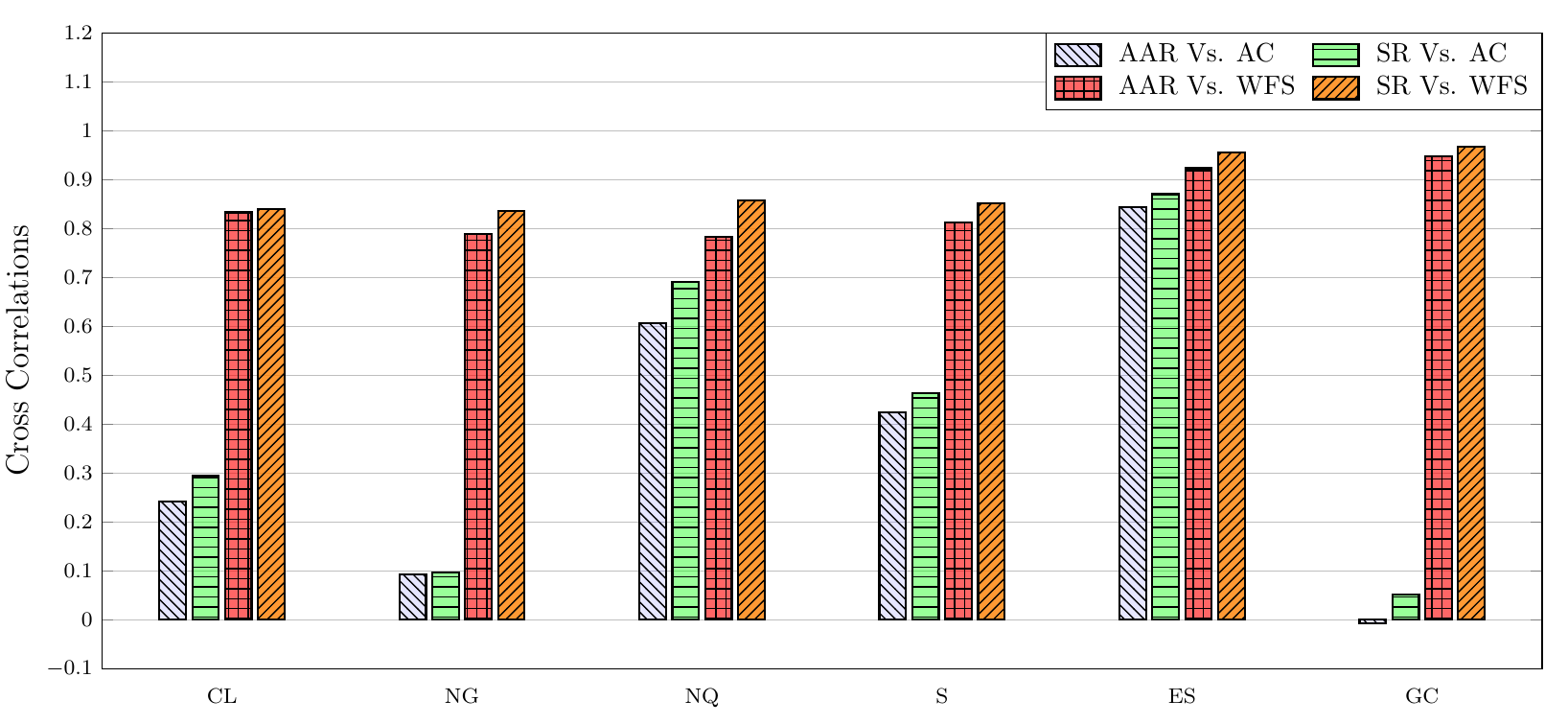}
\caption{\textbf{Cross-Correlation between Metrics in Finance and Machine Learning}. Weighted F score (WFS) highly correlates with average annual return (AAR)/Sharpe ratio (SR) on all the futures' experiments.}
\label{fig:cc}
\end{figure}

To give a whole picture of the feature extraction ability of our CDT 1-D CNN model, we test the following four scenarios: MLP with technical indicators, regular 1-D CNN without technical indicators, CDT 1-D CNN with technical indicators, and CDT 1-D CNN without technical indicators. the average annual return, Sharpe ratio, and weighted F score of the five approaches for the six futures are shown in Figure \ref{fig:ar}, Figure \ref{fig:sr}, and Table \ref{fr}, respectively. The cumulative return and monthly return of the WTI crude oil (CL) futures are plotted in Figure \ref{fig:cm}. We also calculate the cross-correlation between the metrics of finance and machine learning, which is shown in Figure \ref{fig:cc}. Since Nearest-neighbour's results are similar to SVM's, we omit its results in the aforementioned table and figures. 

Our CDT 1-D CNN model achieves the best performance (57.4\% weighted F score, 71.9\% average annual return, and 2.72 Sharpe ratio) for the WTI crude oil (CL) futures. As plotted in Figure \ref{fig:cm}, the cumulative return of the CL futures stays positive during the nearly 7-years period and eventually goes up to 500\%. We notice that our CDT 1-D CNN model is able to turn around quickly when drawdowns happen. Only  {\em two} of the 13 negative return months have lower than 5\% drawdown. In contrast, the highest profitable month has 29.5\% positive return and the profit of 40 months is higher than 5\%. 

Although the regular 1-D CNN model achieves good result sometimes, such as 54.8\% weighted F score (WFS), 10.8\% average annual return (AAR), and 0.76 Sharpe ratio (SR) for the E-mini S\&P 500 (ES) futures, its robustness is not consistent in that its performance for the Natural Gas (NG), Soybeans (S), Gold (GC) futures is as low as that of the baselines. Compared to our CDT 1-D CNN model, the maximum performance degradation of the regular 1-D CNN model is up to 7.0\% for WFS, 40.4\% for AAR and 1.60 for SR for the CL futures, and the average degradation is up to 4.3\% WFS, 25.5\% AAR, and 1.41 SR. These experiments confirm the effectiveness of the scan strategy of our CDT 1-D CNN model.

As plotted in Figure \ref{fig:cc}, the cross-correlation between Accuracy (AC) and average annual return (AAR)/Sharpe ratio (SR) varies from -0.01 for the AAR of the Gold (GC) futures to 0.91 for AAR of the  E-mini S\&P 500 (ES) futures. In contrast, our proposed weighted F score (WFS) highly correlates with both AAR and SR, and has the highest score of cross-correlation for each of the futures. It confirms our argument in Section \ref{Weighted F Scores} that correlation between Weighted F Score and average annual return/Sharpe ratio is significantly larger than that between common metrics in machine learning and average annual return/Sharpe ratio with the most of the futures. 

Technical indicators are effective to MLP, but they bring negative effects to 1-D CNN. As shown in Figure \ref{fig:ar} and \ref{fig:sr}, for commodity futures (CL, NQ, S, and GC),  {\em CDT 1-D CNN with technical indicators} degrade by 6.1\%-53.0\%, 53.0\%-199.0\%, and 2.3\%-6.7\% for average annual return (AAR), Sharpe ratio (SR), and weighted F score (WFS), respectively, compared to that of {\em CDT 1-D CNN without technical indicators}. Except for the probable reason we mentioned in Section \ref{Without technical indicators}, another one is that redundant data types make the parameter sharing difficult to work. Parameters can be updated by back propagation of each data type, but unfortunately, not all technical indicators correlate with price trends all the time. Therefore, some useful features may be counteracted by the unnecessary parameter updates.
\section{Discussion}
\textbf{Long-term features.} Thus far we only consider features within two hours by CDT 1-D convolutions, and relations among data inputs are mutually independent. However, we believe that there may exist informative features within longer time spans. In our future research, We will explore more complex model architectures, such as recurrent neural networks combined with convolutional kernels, in order to capture both short-term and long-term features.

\textbf{Data quality.} We notice that some big losses in our backtest experiments result from data missing in the raw financial trading data. The prices and volume of the current data point may correlate with the upcoming price trend, but may be rarely informative for the one-day-later trend. The data missing of raw financial data also brings negative effect to training process due to the mismatch between the label and the intrinsic information of the current data point. We will develop appropriate data preprocessing methods to solve this issue.

\textbf{Labeling.} Although our experiments verify that supervised training based on our three-class labeling strategy can achieve state-of-the-art performance in both finance and machine learning benchmark, this labeling strategy cannot distinguish more precise classes, such as violent surge, moderate surge, crash, and edge down. We plan to investigate more fine-grained labeling strategy and study its impact on the performance. 
\section{Conclusion}

In this paper, we propose and develop a deep convolutional neural network model to predict the market movement. It uses a novel 1-D convolution, called Cross-Data-Type 1-D Convolution, to capture extracts directly from financial historical trading data. Backtest results show that our model can effectively extract features that are more generalized and informative than those represented by traditional technical indicators. Experiment results show that our model outperforms the baselines by 12.4\%-63.3\% on average annual return and 99\%-245\% on Sharpe ratio over previous machine learning approaches. In addition, we propose a new measure, weighted F-score, which better correlates with financial returns than traditional machine learning performance metrics. 




\end{document}